\documentclass[twocolumn,secnumarabic,amssymb,nobibnotes,aps,prd,showpacs,showkeys]{revtex4-1}

\usepackage{graphicx}
\usepackage{amsmath}
\newcommand{\beq}{\begin{equation}}
\newcommand{\eeq}{\end{equation}}
\newcommand{\eml}{\end{mathletters}}

\newcommand{\be}{\begin{equation}}
\newcommand{\ee}{\end{equation}}
\newcommand{\bea}{\begin{eqnarray}}
\newcommand{\eea}{\end{eqnarray}}
\newcommand{\nn}{\nonumber\\}
\newcommand{\oh}{\frac{1}{2}}

\newcommand{\pa}{\partial}

\begin{document}

\title{Proton emission systematics along proton drip line} 

\author{D.S. Delion$^{1,2,3}$ and A. Pencu$^{1,3}$}
\affiliation{
$^1$Horia Hulubei National Institute for R\&D in Physics and Nuclear Engineering, \\
30 Reactorului, P.O. Box MG-6, RO-077125, Bucharest--M\u agurele, Rom\^ania \\
$^2$Academy of Romanian Scientists, 54 Splaiul Independen\c tei RO-050085,
Bucharest, Rom\^ania \\
$^3$Department of Physics, University of Bucharest,
405 Atomi\c stilor, POB MG-11, RO-077125, Bucharest--M\u agurele, Rom\^ania
}
\date{\today}

\begin{abstract} 
We analyze the chart containing both spontaneous and beta-delayed proton emission processes in terms of 
the Coulomb parameter, reduced radius and angular momentum ($\chi$, $\rho$, $l$).
We then compare the methods to estimate decay width $\Gamma$ of a resonant state in a proton mean field, namely 
the continuity equation for outgoing Gamow states,
phase shift analysis of real scattering states and
numerical integration of the Schr\"odinger equation in the complex plane.
We show that they provide similar results in the region where it is possible to evaluate the 
imaginary part of the energy for a resonant (Gamow) state.
We then investigate the role of the centrifugal barrier induced by Coulomb interaction and also
by proton single particle orbitals. We show that the so-called universal decay law, connecting the logarithm
of the monopole reduced width to the fragmentation potential, remains also valid for beta-delayed proton emission processes.
This fact allows us to describe experimental data for all proton emission processes in terms of a linear dependence 
connecting the logarithm of the monopole Coulomb-reduced decay width to the logarithm of the monopole Coulomb penetrability 
and fragmentation potential within a factor of three for absolute values.
\end{abstract}


\maketitle

\section{Introduction}
\label{sec:intro} 
\setcounter{equation}{0} 
\renewcommand{\theequation}{1.\arabic{equation}} 

Nuclei along proton drip line are very unstable by decaying in different ways.
Many nuclei above doubly magic $^{100}$Sn have one or two protons in continuum which are spontaneously emitted. 
For these emitters the ratio between the Q-value and Coulomb barrier is much smaller than unity. 
They are characterized by relative large half lives ($10^{-6}-10^{3}$ s) and therefore can be described
within the semiclassical approach \cite{Woo97,Son02,Del06a,Pfu12}.
On the other hand, nuclei below $^{100}$Sn decay by $\beta^+$ process leaving valence protons 
in excited resonant states which are rapidly emitted.
This two-step process is called beta-delayed proton emission. The corresponding proton half-lives 
for light nuclei are by only 2-3 orders of magnitude larger than the characteristic nuclear time $10^{-22}$ s, 
while beta-decay half lives being by 10-15 orders of magnitude larger \cite{Cer77,Dos07,Bla08,Lor12,Bat20,ENSDF}.
The excited proton states have Q-values close to the Coulomb barrier, the centrifugal barrier playing an important role
and therefore the description of the beta-delayed proton emission process needs an exact treatment.

Many theoretical descriptions of spontaneous proton emission were proposed within 
the semiclassical approach from spherical \cite{Buc92,Abe97,Bud17,Che20,Oud20} 
or deformed emitters \cite{Mag98,Mag99} in the framework of the time--dependent approach \cite{Tal99,Tal00},
coupled channels method \cite{Bar00,Esb00}, R--matrix theory \cite{Kru04}, 
relativistic mean field theory \cite{Sah11} and also covariant density functional approach \cite{Zha14}.
Nuclear structure was probed by angular proton distribution of emitted protons \cite{Kad00}, 
emission from triaxially deformed nuclei \cite{Dav04}, particle--vibration coupling \cite{Hag01,Dav01}, 
fine structure of transitions to  excited states  \cite{Mag00,Kru00} and from high lying isomers \cite{Liu07}.

It turns out that the logarithm of the decay width corrected by the Coulomb centrifugal barrier
is concentrated along two parallel lines divided by the charge number $Z=68$ \cite{Del06,Med07,Bud17},
or mass number $A=145$, where we notice a strong change from prolate to oblate shapes \cite{Mol95}.
The explanation for this behavior was given in Ref. \cite{Del09} in terms of two regions of the 
fragmentation potential (the difference between the barrier height and proton $Q$--value) and later
by Ref \cite{Del21} in terms of the dependence upon the mass parameter $A^{1/3}$ \cite{Del21}.
An analysis of the proton formation amplitude was also performed in Ref. \cite{Qi12}.

On the other hand, the beta delayed proton decay is a more complex phenomenon, due to the fact 
that it takes place from the excited states created by the primary beta decay process.
In spite of the very large amount of measured Q-values, the availability of experimental proton
half lives is rather limited \cite{Bat20}. A direct estimate of decay widths in beta-delayed proton
emission is performed by analyzing the peaks in the energy spectra in terms of the R-matrix theory \cite{Lan58,War86}.

The purpose of this paper is to describe both spontaneous and beta delayed proton emission within 
a common formalism provided by single particle resonant states, 
allowing to derive a simple Geiger-Nuttal systematics in terms of the monopole Coulomb penetrability
and fragmentation potential. 

\section{Theoretical background}
\label{sec:theor} 
\setcounter{equation}{0} 
\renewcommand{\theequation}{2.\arabic{equation}} 

Spontaneous proton emission along proton drip line with $A>100$ was extensively investigated from theoretical 
point of view \cite{Del06a}. The beta-delayed proton emission is a similar process, but takes place in two steps. 
First $\beta^+$ decay feeds some excited proton state in continuum, which rapidly decays
\bea
A(Z,N) &\rightarrow& B(Z-1,N+1)+e^++\nu
\nn
B(Z-1,N+1) &\rightarrow& C(Z-2,N+1) + p(l,j)~,
\eea
where $(l,j)$ represents the angular momentum and total spin of the emitted proton.
As a distinctive feature, the half life of the primary beta decay process is much larger than
that of the secondary proton emission.
From the nuclear structure point of view there are two main situations:

(i) the nucleus $B$ is odd-odd and the excited states are collective $1^+$ Gammow-Teller modes, 
described by phonons built on proton-neutron pairs $(p\otimes n)_{1^+}$ \cite{Suh06}, as for instance $^{20}$Na \cite{Lun16},

(ii) the nucleus $B$ is odd-mass and the states are built by coupling $1^+$ phonons with an odd proton 
to the total spin $j$, $(1^+\otimes p)_j$ \cite{Del97,Mus06}, as for instance $^{21}$Na \cite{Lun15,Lun15a},

In both situations the proton (of the pair or the odd proton) in a resonant state
of the nuclear mean field penetrates the surrounding Coulomb plus centrifugal barrier. 
Therefore the standard procedure to estimate the decay width is affected only by the 
amplitude of the decaying proton in the collective state.
This resonant state is an eigenstate of the proton mean field potential with complex energy in
daughter nucleus and outgoing asymptotics described by the Coulomb-Hankel function \cite{Del10}.
For outgoing narrow resonances, called Gamow states \cite{Del10,Ber13}, the real part of the energy is positive 
and it is called Q-value.
The decay width $\Gamma$ is twice the complex part of the energy and for narrow resonances it satisfies
the condition $\Gamma<<Q$. Thus, the wave function describing the emission process is given by the following ansatz
\bea
\label{Gamow}
\Psi({\bf R},t)=\Phi({\bf R})\exp\left[-\frac{i}{\hbar}\left(Q-i\frac{\Gamma}{2}\right)t\right]~.
\eea
It corresponds to a pole of the S-matrix in the complex energy plane. 

Proton single particle resonances are usually generated by a Woods--Saxon plus Coulomb potential $V_0({\bf R})$.
Proton wave function in a spherical approach has given spin and parity and can be expressed in terms of
spherical spin--orbit orbital as follows
\bea
\label{Phi}
\Phi_{ljm}({\bf R})=\frac{f_{lj}(R)}{R} \left[Y_{l}(\widehat{R})\otimes\chi_{\oh}({\bf s})\right]_{jm}~,
\eea
where ${\bf R}=(R,\widehat{R})$ denotes the proton--core distance.
At large distances, where the interaction becomes purely Coulombian $V_0(R)\rightarrow V_C(R)=Ze^2/R$,
the radial wave satisfies the standard Coulomb equation
\bea
\label{Couleq}
\left[-\frac{d^2}{d\rho^2}+\frac{l(l+1)}{\rho^2}+\frac{\chi}{\rho}-1\right]f_{lj}(R)=0~.
\eea
Here, the ratio between $V_0(R)$ and $Q$--value acquires the following form
\bea
\frac{V_0(R)}{Q}\rightarrow\frac{V_C(R)}{Q}\equiv\frac{Ze^2}{RQ}=\frac{\chi}{\rho}~,
\eea
in terms of the Coulomb parameter
\bea
\label{chi}
\chi&=&\frac{2Z}{\hbar v}~,~~~v=\sqrt{\frac{2Q}{\mu}}~,
\eea
and reduced radius
\bea
\label{rho}
\rho&=&\kappa R~,~~~\kappa=\frac{\sqrt{2\mu Q}}{\hbar}~,
\eea
where $Z$ is the daughter charge and $\mu$ the reduced mass of the proton--core system.
The outgoing solution of the above equation (\ref{Couleq}), describing an emission process,
is given by the Coulomb-Hankel outgoing wave
\bea 
H^{(+)}_l(\chi,\rho)=G_l(\chi,\rho)+iF_l(\chi,\rho)~,
\eea
depending upon irregular $G_l$ and regular $F_l$ Coulomb waves.
It is related to the monopole function as follows
\bea
\label{Cl}
H^{(+)}_l(\chi,\rho)=C_l(\chi,\rho)H^{(+)}_0(\chi,\rho)~,
\eea
where the coefficient has a closed analytical expression within the semiclassical approach \cite{Del10}.

We use the matching condition at certain radius $R$ where the nuclear interaction vanishes
\bea
f_{lj}(R)=N_{lj}H^{(+)}_l(\chi,\rho)~,
\eea
between the internal radial wave function normalised to unity inside nucleus 
and external Coulomb-Hankel outgoing wave.
The matching coefficient $N_{lj}$, called scattering amplitude, does not depend upon the radius 
at relative large distances because both functions satisfy the same Schr\"odinger equation.
We estimate the decay width by using complex energy $E=Q-i\Gamma/2$ in the stationary Schr\"odinger
equation and its complex conjugate \cite{Del10}
\bea
\label{Gamma1}
\Gamma^{(1)}_{lj}=\hbar v |N_{lj}|^2~.
\eea
This relation can be rewritten as a standard factorisation \cite{Lan58}
\bea
\label{fact}
\Gamma^{(1)}_{lj}=P_l\gamma_{lj}^2~,
\eea
in terms of Coulomb penetrability and reduced width
\bea
\label{factor}
P_l=\frac{2\rho}{\left|H^{(+)}_l(\chi,\rho)\right|^2}~,~~~
\gamma_{lj}^2=\frac{\hbar^2f_{lj}^2(R)}{2\mu R}~.
\eea
In the alternative description, by using real scattering states with given angular momentum and energy
\bea
\label{asymp}
&&f_{lj}(R,E)\sim G_l(\chi,\rho)\sin\delta_{lj}(E)+F_l(\chi,\rho)\cos_l\delta_{lj}(E)
\nn
&=&\frac{i}{2}e^{-i\delta_{lj}(E)}\left[H^{(-)}_l(\chi,\rho)-S_{lj}(E)H^{(+)}_l(\chi,\rho)\right]~,
\eea
defining the S-matrix $S_{lj}(E)=\exp[2i\delta_{lj}(E)]$, the phase shift for some narrow resonant state 
sharply passes through the value $\delta_{lj}=\pi/2$ where only the irregular component $G_l$ remains.
At this value the decay width is estimated according to the following ansatz \cite{Del10}
\bea
\label{Gamma2}
\Gamma^{(2)}_{lj}=-2\left[\frac{\pa\cot\delta_{lj}(E)}{\pa E}\right]^{-1}_{E=Q}~.
\eea
For a relative narrow resonance, where one has $|G_l(\chi,\rho)|>>|F_l(\chi,\rho)|$ inside the barrier \cite{Del10}, 
both complex Gamow and real scattering state descriptions give close results.

The proton is born usually from a collective state with a probability $p_{lj}$ (spectroscopic factor) and 
multiplicity $\Omega_j=j+\oh$. Therefore the decay width acquires the following ansatz \cite{Del10,Del21}
\bea
\Gamma_{lj}^{(k)}\rightarrow \frac{p_{lj}}{\Omega_j}\Gamma_{lj}^{(k)}~,~~~k=1,2,3~.
\eea
Here, the index $k=3$ corresponds the outgoing Gamow resonant state (\ref{Gamow}), found by directly solving
Schr\"odinger equation in the complex plane \cite{Ver82,Ixa95}.

The spectroscopic factor for spontaneous emission is given by the BCS (Bardeen-Cooper-Schrieffer) 
particle probability $p_{lj}=u^2_{lj}$ in the quasiparticle representation. 
One has a value $u^2_{lj}\sim 0.5$ for a state close to the Fermi level, which rapidly approaches unity for higher states.
For odd-odd nuclei described within pn-QRPA formalism \cite{Suh06}, the spectroscopic factor is described by
the forward amplitude of the Gamow-Teller $1^+$ n-th resonance $p_{lj}=X^2_{lj}(n)$.
It turns out that, except for the lowest eigenstate $n=1$, the higher eigenstates of the $1^+$ phonon
are rather pure, by containing practically one quasiparticle proton and neutron states.
Similar conclusion holds for the triplet representation in odd-mass nuclei described within 
the phonon-quasiparticle coupling \cite{Del97,Mus06}.
Let us also mention that the deformation correction gives an increase of the decay width by about 30\%
for a quadrupole deformation $\beta_2=0.3$ \cite{Del21}.
Therefore the order of magnitude of the decay width is not strongly affected by the nuclear structure details.

\section{Proton emission systematics}
\label{sec:numer} 
\setcounter{equation}{0} 
\renewcommand{\theequation}{3.\arabic{equation}} 

\subsection{Experimental data}

We will split emitters on the proton drip line according to Table I.
At present we have at our disposal a relative large amount of experimental data for proton emitters
with $A>100$ \cite{Del21}. They are split into first two zones in Table I corresponding to
two main regions of the fragmentation potential $V_{frag}=V_B-Q$ separated by $Z=68$, or $A=145$
\cite{Del06,Del21}. One the other hand, there is a rather limited amount of measured beta-delayed 
proton decays widths, given by zone 3, containing $^{20-21}$Na isotopes \cite{Lun16,Lun15,Lun15a}. 
Zone 4 with $22<A<100$ contains proton-delayed proton emitters with only Q-values measured \cite{Bat20}

	\begin{table}[ht]
		\centering
		\caption{Zones of proton emitters}
\begin{tabular}{|c|c|c|c|}
	\hline
	Zone & $Z$ & $A$ & Symbol \\
	\hline
	1  & 50-68 & 100-145 & dark circles \\
	2  & $>$68 & $>$145  & open circles \\
	3  & 11-11 & 20-21   & dark squares \\
	4  & 12-49 & 22-100  & open squares \\
	\hline
\end{tabular}
	\end{table}

Proton emission data from the odd-odd emitter $^{20}$Na$\rightarrow^{19}$Ne+p 
(angular momentum, Q-value, decay width, Coulomb-reduced decay width, reduced width, $\rho$,  $\chi$, 
monopole penetrability, Coulomb reduction coefficient and fragmentation potential) are given in Table II \cite{Lun16}.

\begin{figure}[ht] 
\begin{center} 
\includegraphics[width=9cm]{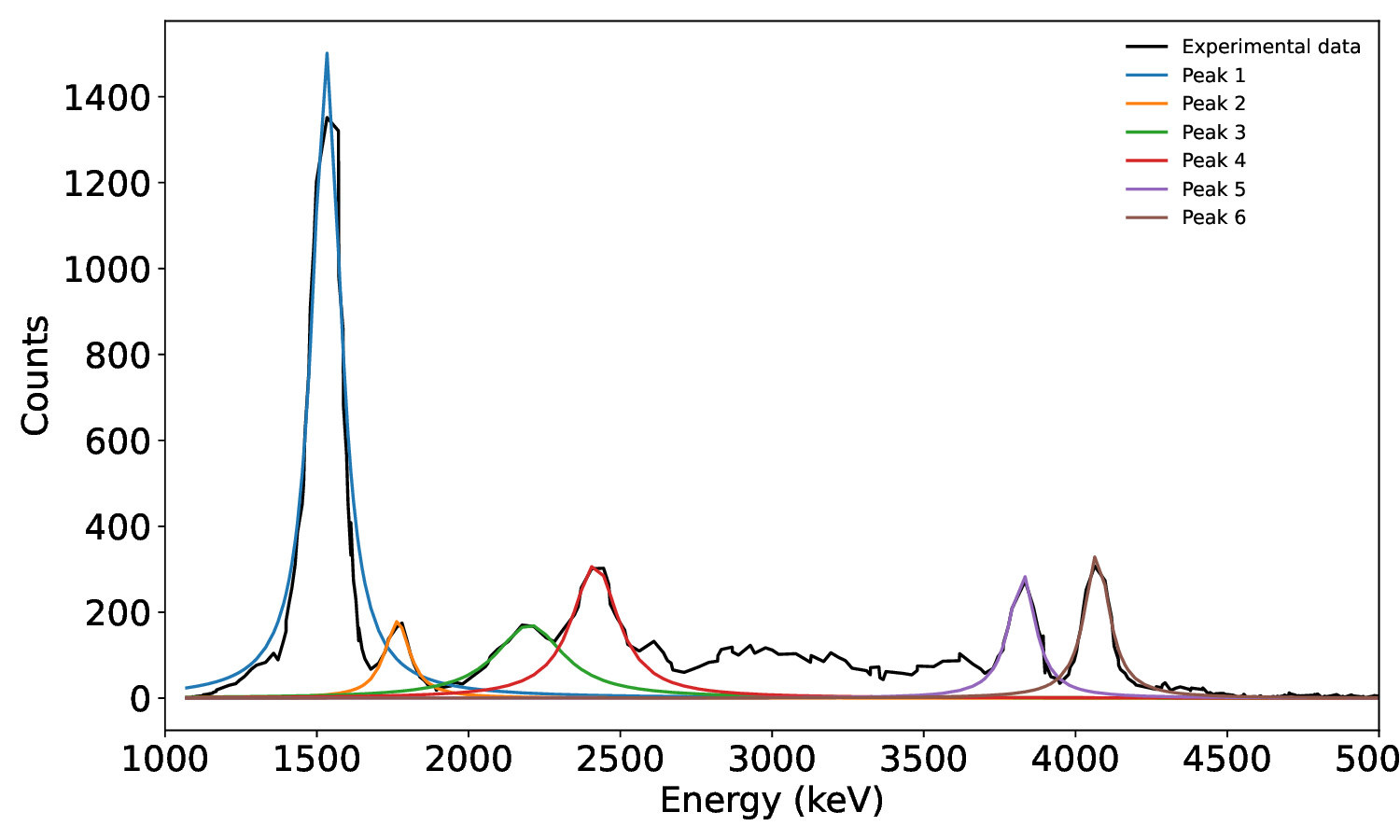} 
\vspace{-5mm}
\caption{
Fit of the proton experimental spectrum with Lorentzian dependencies in $^{20}$Na.
}
\label{fig1}
\end{center} 
\end{figure}

In order to approximate the beta delayed proton emission decay widths for $^{20}$Na we have digitized 
the experimental data provided in Ref. \cite{Lun16} and performed a fit of the peaks using Lorentz distributions, 
according to the R-matrix standard procedure \cite{Lan58}. The scale parameter of each distribution is then proportional 
to the decay width of the respective channel. Using the exponential centrifugal factor (\ref{Cl}) we can differentiate 
between close proton Q values with different angular momenta.
In Fig. \ref{fig1} we give an example af several Lorentzian distributions fitting the experimental spectrum.

Similar data for emission from the odd-mass emitter $^{20}$Na$\rightarrow^{19}$Ne+p are given in Table III,
where both Q-values and decay widths are taken from Refs. \cite{Lun15,Lun15a}.

In beta-delayed proton emission from excited states the Q-value lies much closer to the top of the Coulomb+nuclear barrier
compared to spontaneous emission. The actual radius and value of the barrier is given
by the following linear dependence upon the mass number
\bea
R_B=(0.93\pm 0.06)A^{1/3}+(4.65\pm 0.25)~.
\eea
As a result, the value of the Coulomb+nuclear barrier at this radius becomes by 7\% smaller than the corresponding pure Coulomb barrier
\bea
\label{VB}
V_B=0.93V_C(R_B)=0.93\frac{Ze^2}{R_B}~.
\eea

\begin{widetext}
\begin{center}
	\begin{table}[ht]
		\centering
		\caption{Proton emission data for $^{20}$Na$\rightarrow^{19}$Ne+p}
\begin{tabular}{|c|c|c|c|c|c|c|c|c|c|c|}
	\hline
	No& l & $Q$ (MeV) & $\Gamma_l$ (keV)& $\Gamma_{red}$ (keV)& $\log_{10}{\gamma_0^2}$ & $\rho$ & $\chi$ & $P_l$ & $C_l^2$ & $V_{frag}$ (MeV) \\
	\hline
	1 & 1 & 1.622 & 14.980 & 43.730 & -1.008 & 1.200 & 2.420 & 0.153 & 2.919 & 1.976 \\
	2 & 0 & 1.656 & 43.730 & 43.730 & -1.032 & 1.213 & 2.395 & 0.471 & 1.000 & 1.942 \\
	3 & 0 & 1.853 & 21.190 & 21.190 & -1.468 & 1.283 & 2.264 & 0.623 & 1.000 & 1.745 \\
	4 & 1 & 1.905 & 8.259 & 21.190 & -1.496 & 1.300 & 2.233 & 0.259 & 2.565 & 1.693 \\
	5 & 0 & 1.907 & 21.190 & 21.190 & -1.497 & 1.301 & 2.232 & 0.666 & 1.000 & 1.691 \\
	6 & 1 & 2.320 & 116.700 & 255.900 & -0.592 & 1.435 & 2.024 & 0.455 & 2.194 & 1.278 \\
	7 & 2 & 2.344 & 21.040 & 255.900 & -0.600 & 1.443 & 2.013 & 0.084 & 12.160 & 1.254 \\
	8 & 0 & 2.560 & 37.130 & 37.130 & -1.506 & 1.508 & 1.926 & 1.191 & 1.000 & 1.038 \\
	9 & 0 & 2.567 & 37.130 & 37.130 & -1.508 & 1.510 & 1.924 & 1.197 & 1.000 & 1.031 \\
	10 & 1 & 2.620 & 18.550 & 37.130 & -1.523 & 1.525 & 1.904 & 0.619 & 2.002 & 0.978 \\
	11 & 1 & 4.033 & 18.340 & 28.070 & -1.903 & 1.892 & 1.535 & 1.466 & 1.530 & -0.435 \\
	12 & 0 & 4.051 & 28.070 & 28.070 & -1.905 & 1.896 & 1.531 & 2.255 & 1.000 & -0.453 \\
	13 & 2 & 4.053 & 6.295 & 28.070 & -1.905 & 1.897 & 1.531 & 0.506 & 4.458 & -0.455 \\
	14 & 0 & 4.303 & 40.540 & 40.540 & -1.774 & 1.955 & 1.486 & 2.412 & 1.000 & -0.705 \\
	15 & 2 & 4.347 & 10.160 & 40.540 & -1.779 & 1.964 & 1.478 & 0.611 & 3.992 & -0.749 \\
	\hline
\end{tabular}
	\end{table}
		
	\begin{table}[ht]
		\centering
		\caption{Proton emission data for $^{21}$Na$\rightarrow^{20}$Ne+p}
\begin{tabular}{|c|c|c|c|c|c|c|c|c|c|c|}
	\hline
	No& l & $Q$ (MeV) & $\Gamma_l$ (keV)& $\Gamma_{red}$ (keV)& $\log_{10}{\gamma_0^2}$ & $\rho$ & $\chi$ & $P_l$ & $C_l^2$ & $V_{frag}$ (MeV) \\
	\hline
	1 & 0 & 0.402 & 21.000 & 21.000 & 1.624 & 0.606 & 4.867 & 4.99$~10^{-4}$ & 1.000 & 3.152 \\
	2 & 2 & 1.112 & 0.016 & 0.696 & -2.295 & 1.007 & 2.927 & 3.05$~10^{-3}$ & 44.920 & 2.442 \\
	3 & 2 & 1.862 & 3.930 & 71.870 & -0.955 & 1.303 & 2.262 & 0.035 & 18.290 & 1.692 \\
	4 & 2 & 2.036 & 21.000 & 323.000 & -0.389 & 1.363 & 2.163 & 0.051 & 15.380 & 1.518 \\
	5 & 2 & 2.588 & 136.000 & 1310.000 & 0.022 & 1.537 & 1.918 & 0.129 & 9.630 & 0.966 \\
	6 & 0 & 3.543 & 235.000 & 235.000 & -0.921 & 1.798 & 1.640 & 1.962 & 1.000 & 0.011 \\
	7 & 2 & 4.036 & 173.000 & 750.200 & -0.485 & 1.919 & 1.536 & 0.528 & 4.337 & -0.482 \\
	8 & 2 & 5.177 & 204.000 & 615.200 & -0.683 & 2.173 & 1.356 & 0.984 & 3.016 & -1.623 \\
	9 & 2 & 6.543 & 0.650 & 1.490 & -3.390 & 2.443 & 1.206 & 1.596 & 2.292 & -2.989 \\
	\hline
\end{tabular}
	\end{table}
\end{center}
\end{widetext}

\begin{figure}[ht] 
\begin{center} 
\includegraphics[width=9cm]{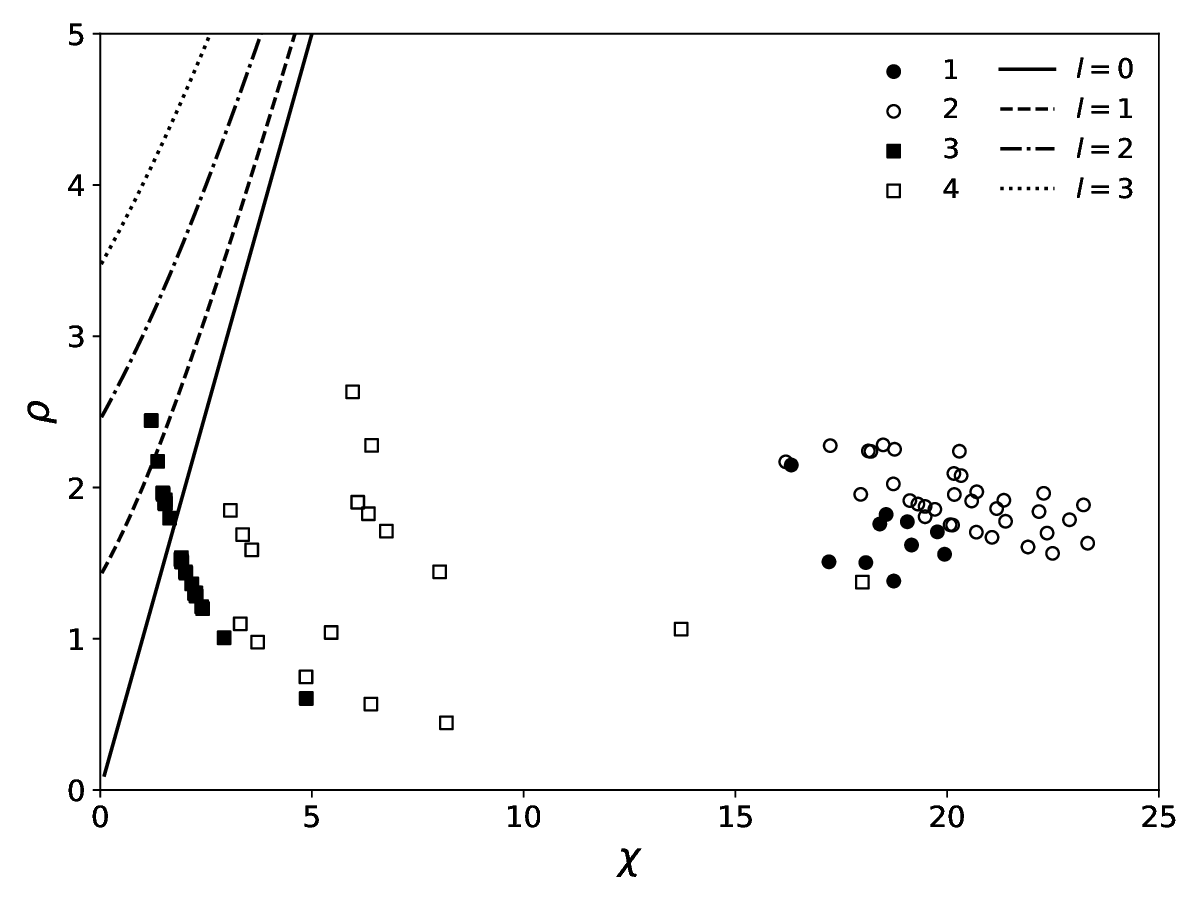} 
\vspace{-5mm}
\caption{
Dependence of the reduced radius $\rho$ estimated at the barrier (\ref{VB}) versus Coulomb parameter $\chi$.
The critical curves corresponding to different angular momenta are solutions of the Coulomb+centrifugal equation (\ref{crit}). 
Circles on the right hand side of the figure correspond to proton emitters with $A>100$ \cite{Del21},
while dark squares on the left hand side correspond to $^{20-21}$Ne and open squares to some representative emitters 
with known Q-values $^{13}$Al, $^{18}$Ar, $^{30}$Zn, $^{69}$Kr and $^{73}$Sr, taken from Ref. \cite{Bat20},
as described by Table I.
}
\label{fig2}
\end{center} 
\end{figure}

\begin{figure}[ht] 
\begin{center} 
\includegraphics[width=9cm]{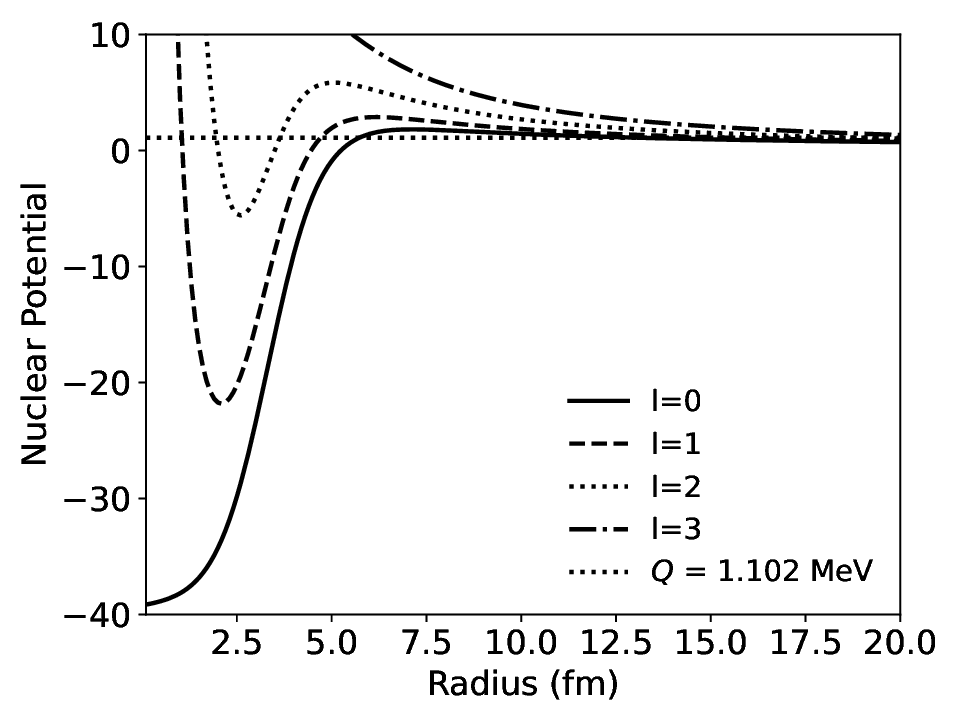} 
\vspace{-5mm}
\caption{
Woods-Saxon nuclear+Coulomb+centrifugal potential with universal parametrisation versus radius for $^{21}$Ne 
corresponding to $l$=0, 1, 2, 3. The horizontal dashed line denotes Q-value.
}
\label{fig3}
\end{center} 
\end{figure}

In order to compare spontaneous with beta-delayed proton emission processes
we plotted in Fig. \ref{fig2} the dependence of the reduced radius $\rho$ estimated at the 
above barrier (\ref{VB}) versus Coulomb parameter $\chi$. Various lines correspond to the critical curves
\bea
\label{crit}
\frac{l(l+1)}{\rho^2}+\frac{\chi}{\rho}-1=0~,
\eea
dividing the sub-barrier (below) and over-barrier (above) regions for each angular momentum $l$.
For instance $\chi/\rho>1$ corresponds to the sub-barrier region with $l=0$. 
Various symbols correspond to the regions in Table I.
The right cluster of values (circles) corresponds to spontaneous proton emission data.
Notice here the two regions divided by $Z=68$ corresponding to different areas of the fragmentation potential
mentioned by Ref. \cite{Del06}. 
The left cluster corresponds to $^{20-21}$Na isotopes with measured decay widths and Q-values (dark squares)
and some representative beta-delayed proton emitters with known Q-values
 $^{13}$Al, $^{18}$Ar, $^{30}$Zn, $^{69}$Kr and $^{73}$Sr, given by Ref. \cite{Bat20} (open squares). 
The Ne cluster (zone 3) is obviously much closer to the top of the Coulomb plus centrifugal barrier 
and therefore correspond to much shorter half-lives.

An example of the nuclear+Coulomb+centrifugal barrier for angular momenta $l$=0, 1, 2, 3
is plotted in Fig. \ref{fig3} for $^{21}$Ne. Here we used Woods-Saxon interaction with universal
parametrisation \cite{Dud82,Cwi87}. 
By a horizontal dashed line it is shown the Q-value of the first resonant state.
Here we notice the important role played by the centrifugal term, significantly enhancing
the total barrier and therefore of the half-life of the resonant state.

\begin{figure}[ht] 
\begin{center} 
\includegraphics[width=9cm]{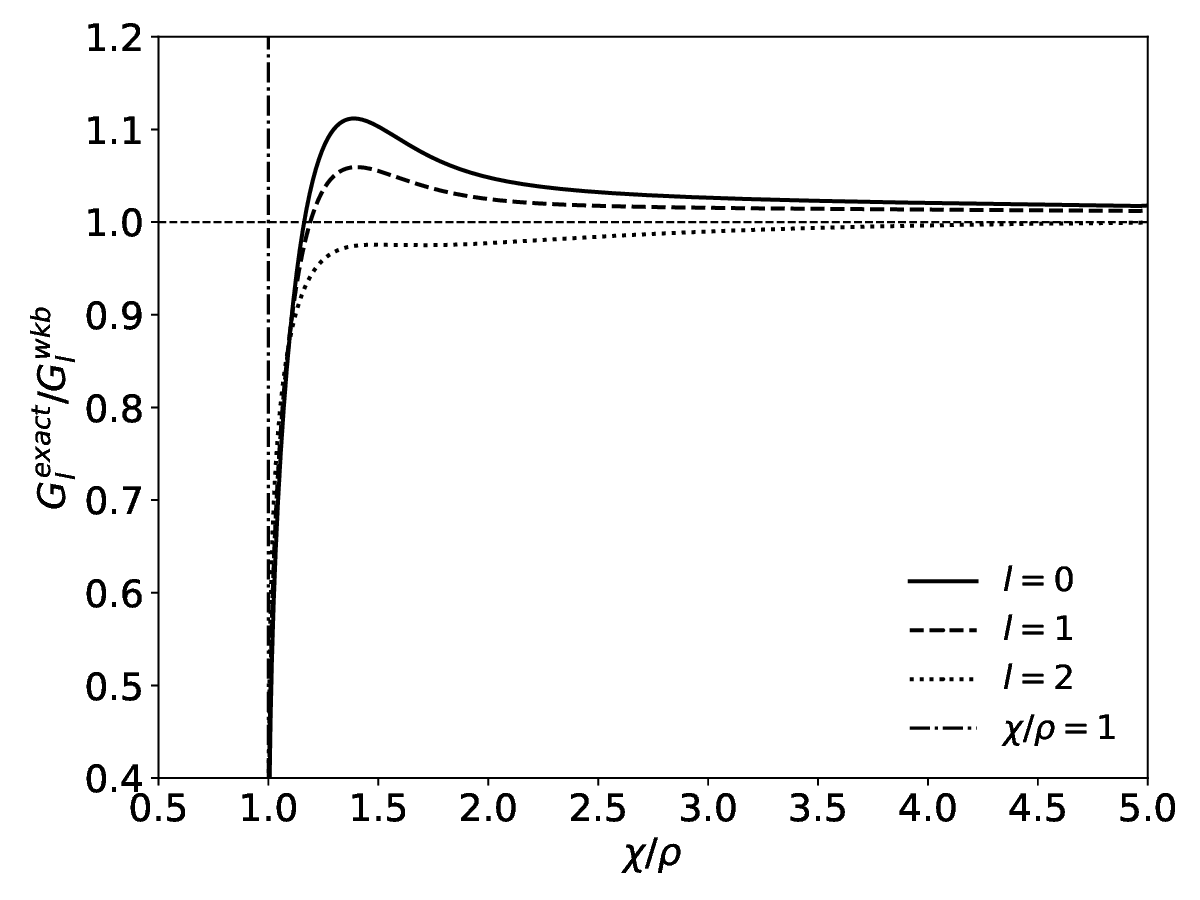} 
\vspace{-5mm}
\caption{
Ratio between exact and WKB irregular function $G_l(\chi,\rho)$ versus $\chi/\rho$ for l=0, 1 , 2.
}
\label{fig4}
\end{center} 
\end{figure}

\subsection{Data analysis}

In applying the relation (\ref{Gamma1}) to estimate the spontaneous decay width from ground state one usually 
uses the WKB (Wentzel-–Kramers–Brillouin) semiclassical analytical relation to estimate Coulomb functions, 
due to the fact that ratio $\chi/\rho>>1$. 
This is true for the right cluster of circles in Fig. \ref{fig2}. For beta-delayed proton emissions, corresponding 
to most emitters of the left cluster in Fig. \ref{fig2}, this condition is not satisfied anymore,
as can be seen from Fig. \ref{fig4}, where we compared exact and WKB irregular Coulomb waves with $l=0,~1,~2$
as a function of the ratio $\chi/\rho$, entering Coulomb equation (\ref{Couleq}). 
For this reason, especially in the region with $\chi/\rho <$ 3, it is necessary to use the exact expressions.

\begin{figure}[ht] 
\begin{center} 
\includegraphics[width=9cm]{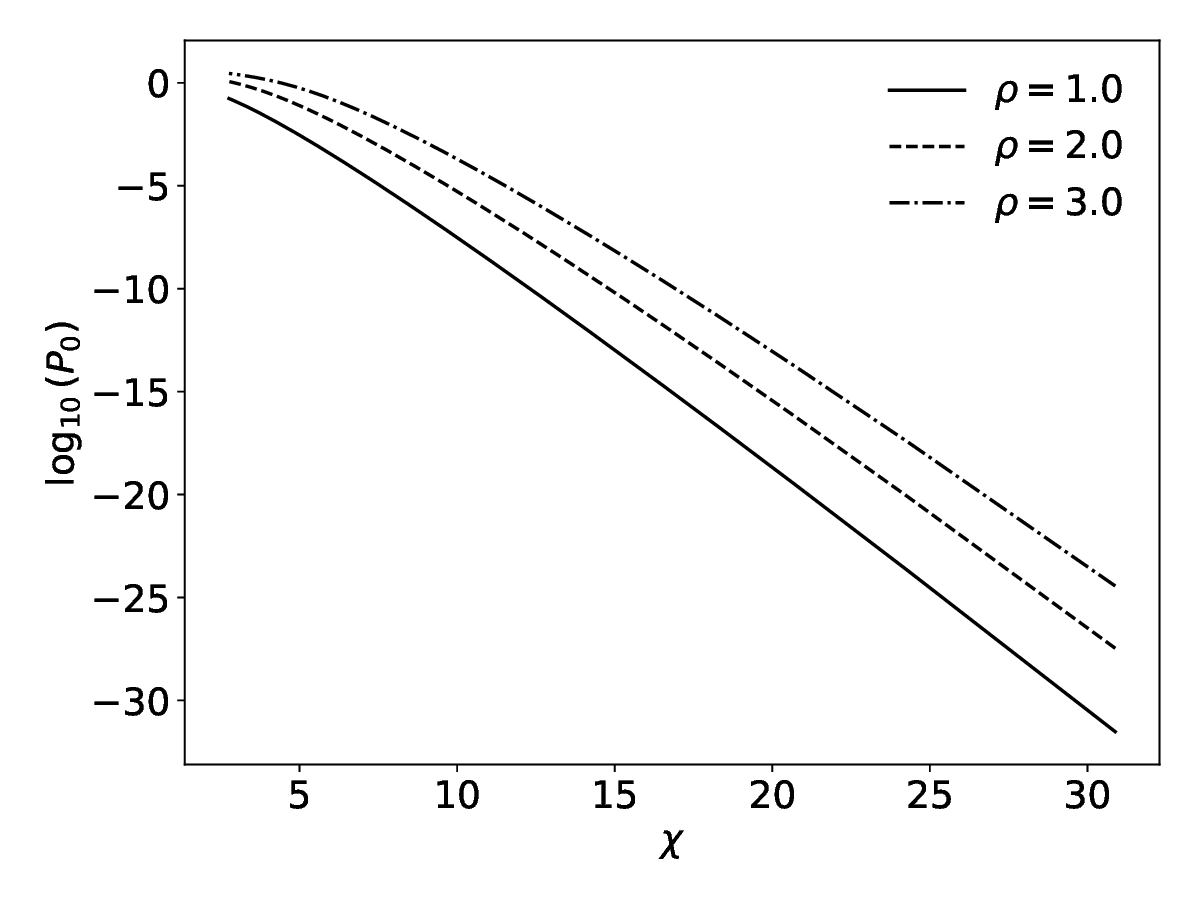} 
\vspace{-5mm}
\caption{
Logarithm of the monopole Coulomb penetrability $P_0(\chi,\rho)$ versus Coulomb parameter $\chi$ for $\rho$= 1, 2, 3.
}
\label{fig5}
\end{center} 
\end{figure}

It is known that the Geiger-Nuttal law linearly connects the logarithm of the decay width (or half life)
to the Coulomb parameter $\chi$. This is based on the representation (\ref{fact}) of the decay width which becomes
\bea
\log_{10}\Gamma_{lj}=\log_{10}P_l+\log_{10}\gamma^2_{lj}~,
\eea
and the proportionality $\log_{10}P_l\sim\chi$, valid for the WKB representation of the Coulomb wave.
From Fig. \ref{fig5}, where we plotted the logarithm of Coulomb penetrability versus Coulomb parameter $\chi$,
we notice that this proportionality fails for $\chi<$ 7, i.e. in the region of beta-delayed  proton emission in Fig. \ref{fig2}.
For this reason we will give our results in terms of the generalized Geige-Nuttal law connecting
the logarithm of the decay width (or half life) to the Coulomb penetrability.

\begin{figure}[ht] 
\begin{center} 
\includegraphics[width=9cm]{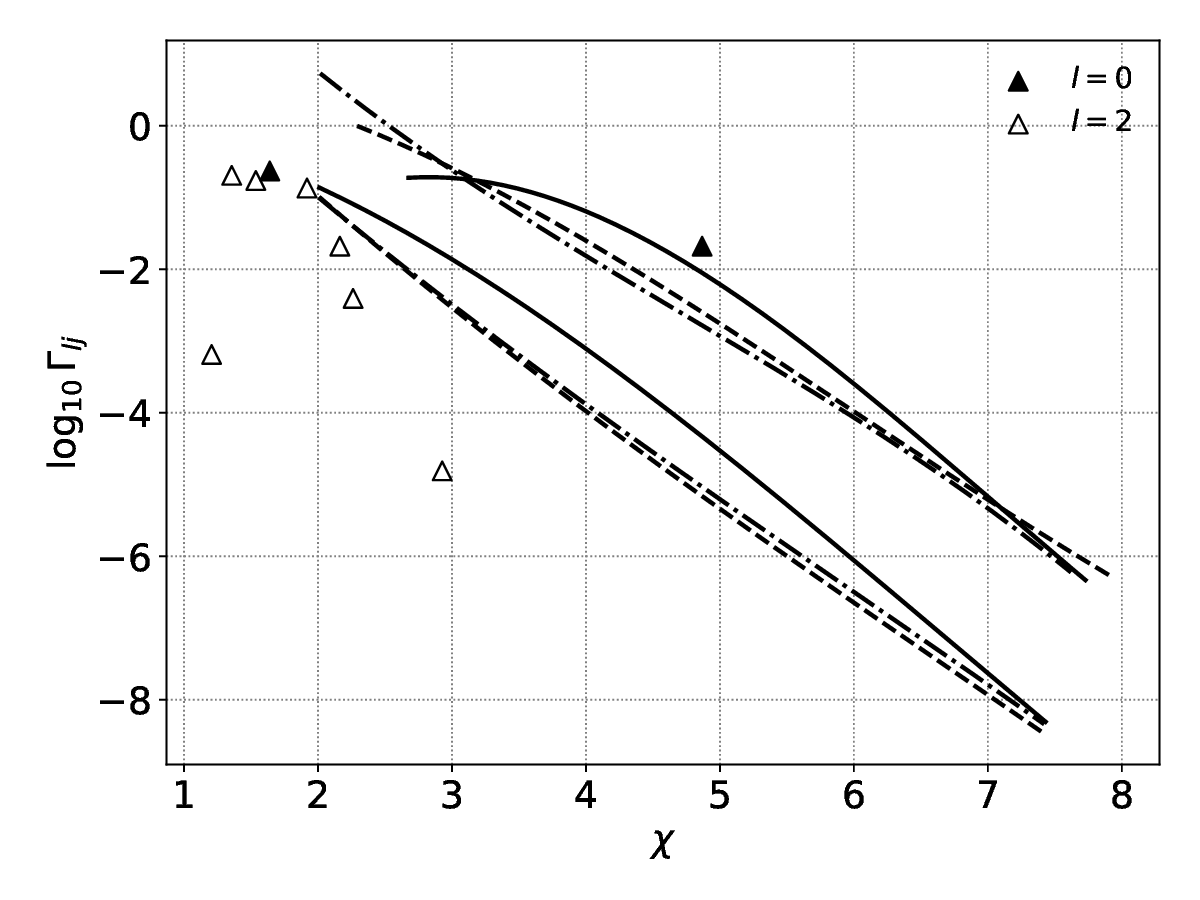}
\vspace{-5mm}
\caption{
Logarithm of the decay width provided by resonant states with $l=0$ (upper curves) and $l=2$ (lower curves)
in the mean field with universal parametrisation versus Coulomb parameter $\chi$ for $^{21}$Na. 
Each line corresponds to $\Gamma^{(k)}_l$ estimated by using different methods k=1 (\ref{Gamma1}) (dashes), k=2 (\ref{Gamma2})
(dot-dashes) and k=3: complex pole in the energy plane (solid line). 
Symbols denote available experimental data.
}
\label{fig6}
\end{center} 
\end{figure}

The properties of the mean field along the proton drip line were analyzed by several papers, see \cite{Ham96},
and predictions were performed by using Hartree-Fock-Bogoljubov approach with Gogny interaction \cite{pot}.
It is not our purpose to obtain the parameters of the proton mean field giving the best description
of experimental decay widths, especially due to the fact that most emitters are deformed.
Anyway, we can understand many gross features of the proton emission process by
performing a theoretical simulation of resonant eigenstates with a given angular momentum and spin $(l,j)$, 
generated by a central plus Coulomb spherical mean field, with the so--called universal parametrisation \cite{Dud82,Cwi87}. 
In Fig. \ref{fig6} we perform a comparison between the above described three methods to estimate decay width $\Gamma^{(k)}_l$,
k=1 (\ref{Gamma1}), (dashes), k=2 (\ref{Gamma2}) (dot-dashes) and k=3, complex pole in the energy plane (solid line).
In order to obtain the relation between decay width and Coulomb parameter $\chi$ we changed the strength of the 
central Woods-Saxon mean field for $^{21}$Ne. We investigated the eigenstates with angular momenta $l=0$ (upper curves) 
and $l=2$ (lower curves).
We notice that the first two methods, k=1, 2, give very close results even for small values of $\chi$, 
while the direct integration in the complex energy plane k=3 deviates from them in this interval, 
but it approaches them for $\log_{10}\Gamma < $ -7. In spite of the fact that we used the mean field with 
universal parametrisation. the curves follow the trend of experimental data plotted by dark triangles ($l=0$)
and open triangles ($l=2$).

In Ref. \cite{Del06} we introduced the monopole Coulomb-reduced decay width defined by the following relation
\bea
\label{red}
\Gamma_{red}=\Gamma_{lj}\vert C_l(\chi,\rho) \vert^2~,
\eea
where the coefficient expressing the l-th Coulomb wave in terms of the monopole component is given by (\ref{Cl})
estimated at the experimental Q-value and nuclear surface $R_0=1.2A^{1/3}$.
In this way the influence of the Coulomb barrier is reduced to a monopole component. 
This procedure was applied to spontaneous proton emission by reducing several parallel lines
of experimental data, corresponding to different angular momenta, 
to only two main monopole Coulomb-reduced lines \cite{Del06}.

\begin{figure}[ht] 
\begin{center} 
\includegraphics[width=9cm]{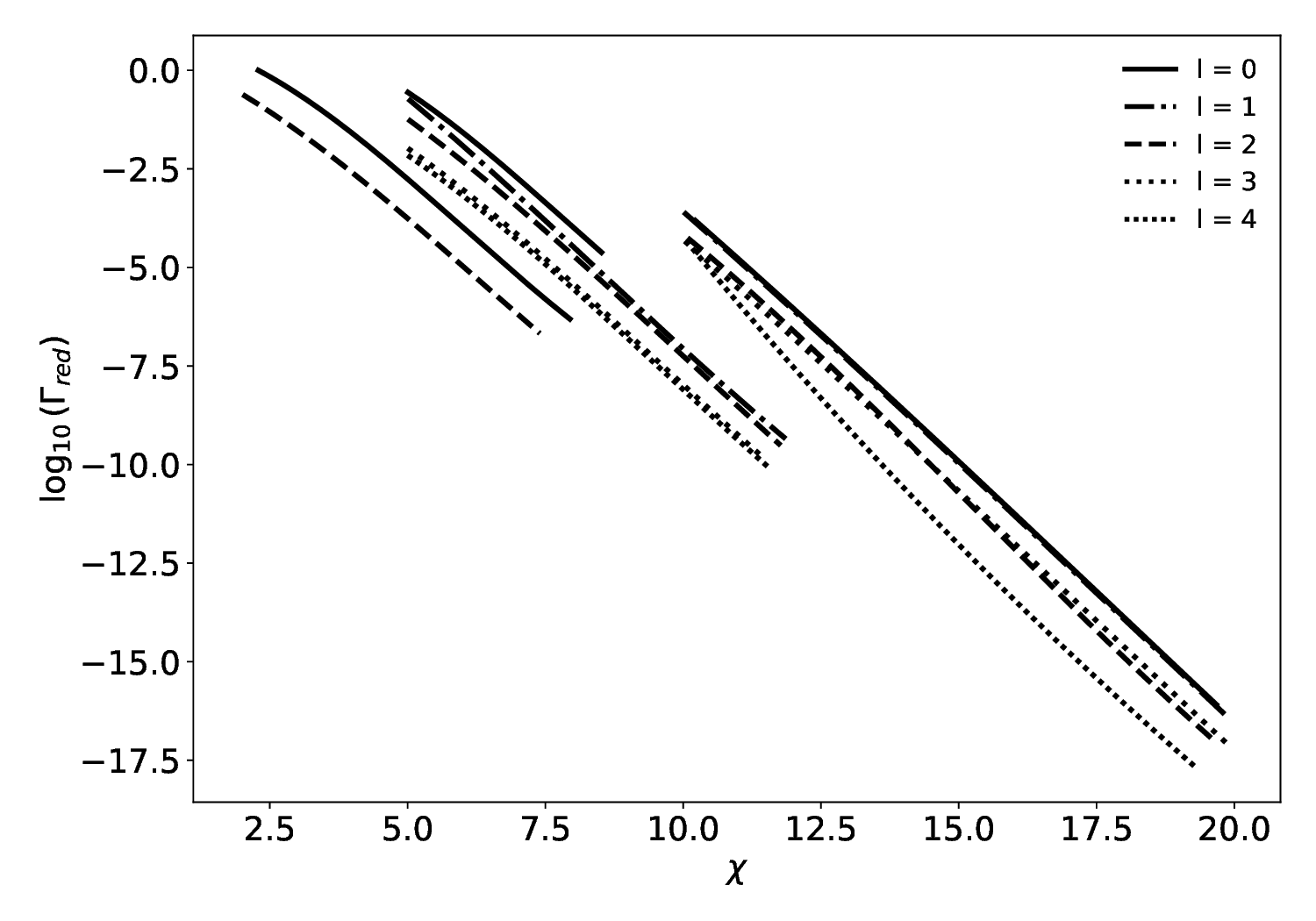}
\vspace{-5mm}
\caption{
Logarithm of the monopole Coulomb-reduced decay width (\ref{Gamma1}) 
provided by resonant states in the mean field with universal parametrisation
versus the Coulomb parameter $\chi$ 
for $^{21}$Na (left group), $^{59}$Zn (middle group) and $^{131}$Eu (right group)
corresponding to angular momenta $l=0,~1,~2,~3,~4$.
}
\label{fig7}
\end{center} 
\end{figure}

We are interested to investigate in a similar way the widths for both spontaneous emission, 
as well as for beta delayed proton decays. First of all in Fig. \ref{fig7} we analyzed the monopole 
Coulomb-reduced decay width according to Eq. (\ref{red}). 
We used three isotopes corresponding to different mass numbers, covering the area from beta delayed
to spontaneous proton emission processes, namely $^{21}$Ne, $^{59}$Zn and $^{131}$Eu,
by Coulomb-reducing data for different angular momenta $l=0,~1,~2,~3,~4$ plotted by using different types of lines. 
As usually, we increased the Q-value of the resonant state by changing the strength of the central potential 
with universal parametrisation.
We notice the linear dependence between $\log\Gamma$ and $\chi$, characterizing the Geiger-Nuttall law, 
except for small values of the Coulomb parameter, where log penetrability looses the linear dependence
upon $\chi$, as given by Fig. \ref{fig5}.
Then we also notice that, after the Coulomb angular momentum reduction, there are three main clusters of lines 
corresponding these isotopes, but inside each cluster still remains a system of parallel curves, 
corresponding to different angular momenta. It turns out that the intercepts of the straight lines
fitting the three clusters are linearly decreasing upon the mass parameter $A^{1/3}$, as we already 
noticed in Ref. \cite{Del21}.

\begin{figure}[ht] 
\begin{center} 
\includegraphics[width=9cm]{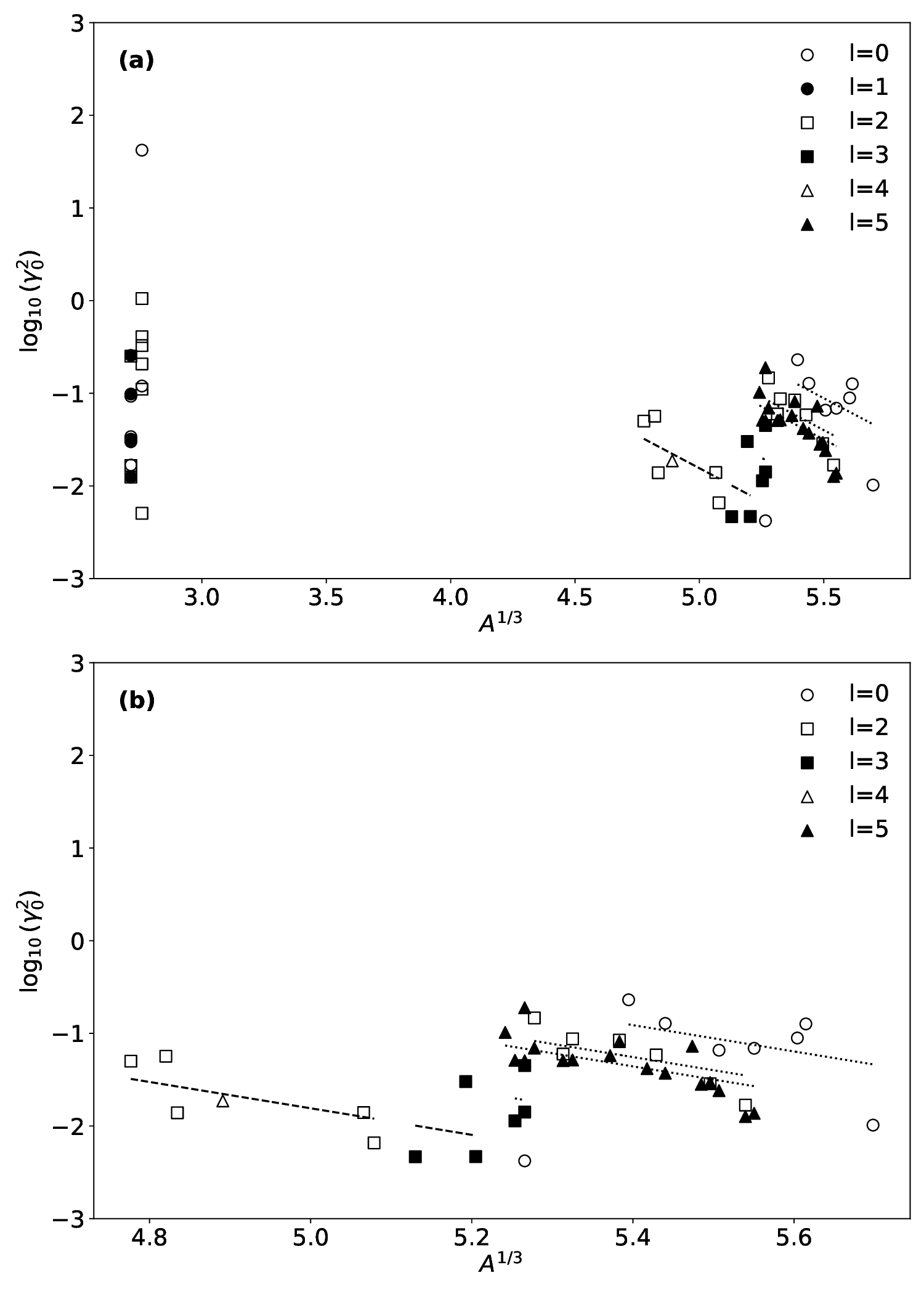}
\vspace{-5mm}
\caption{
Logarithm of the reduced width versus the mass parameter $A^{1/3}$ for all emitters (a)
and for emitters with $A>100$ (b). Symbols denote emitters with different angular momenta.
The corresponding parallel lines with a common slope given by Eq. (\ref{estim}) fit data with different angular momenta.
}
\label{fig8}
\end{center} 
\end{figure}

\begin{figure}[ht] 
\begin{center} 
\includegraphics[width=9cm]{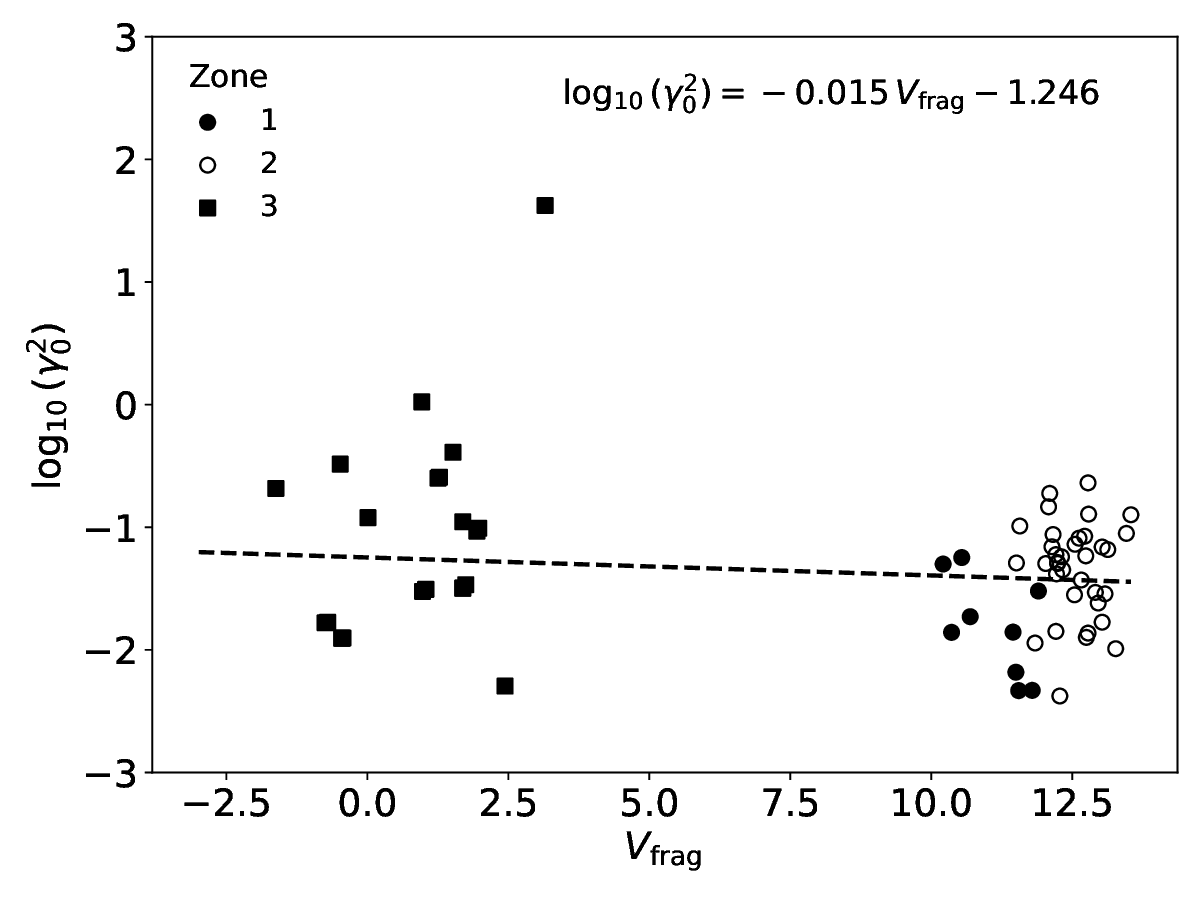}
\vspace{-5mm}
\caption{
Logarithm of the reduced width versus the fragmentation potential $V_{frag}=V_B-Q$. 
Dark circles denote emitters with $100<A<145$ (zone 1)
and dark squares emitters with $A=20,21$ (zone 3).
}
\label{fig9}
\end{center} 
\end{figure}

We can explain the additional splitting of lines among the three clusters by analysing the overall 
decreasing behavior as a ``bulk'' nuclear structure effect induced by the well known asymptotic behavior 
of the proton wave function on the nuclear surface $R_0=1.2A^{1/3}$ \cite{Del10}. 
By taking into account the strongest exponential dependence of the nuclear orbital 
versus angular momentum and mass number
\bea
f_{lj}(R)&\sim& R^{l+1}\exp\left[-\oh\left(\frac{R}{b}\right)^2\right]
\nn
b^{-2}&=&\frac{m_p\omega_0}{\hbar}=0.024\hbar\omega_0~,
\eea
one obtains for the reduced width at the nuclear surface \cite{Del10} by using $\hbar\omega_0=41A^{-1/3}$
\bea
\label{estim}
&&\log_{10}\gamma^2_l(R_0)=\log_{10}\frac{\hbar^2f_{lj}^2(R_0)}{2\mu R_0}
\nn
&\sim&(2l+1)\log_{10}(1.2A^{1/3})-1.42A^{1/3}~.
\eea
Fig. \ref{fig8} confirms this theoretical result. In panel (a) we plotted the logarithm of the 
experimental reduced width 
\bea
\label{exp}
\gamma^2_0=\frac{\Gamma_{red}^{(exp)}}{P_0}~,
\eea
versus $A^{1/3}$ for all experimental data in both beta-delayed decay
region (left zones 3 an 4) and spontaneous emission (right zones 1 and 2). 
In the last region, plotted in panel (b), we notice the system of parallel lines corresponding to different 
angular momenta with a common slope predicted by the above dependence (\ref{estim}). 
Notice that the intercepts of lines corresponding to even angular momenta follow a decreasing pattern, 
while the intercepts of lines with odd angular momenta, corresponding to intruder negative parity orbitals
in this N=4 major shell, follow an opposite trend.
Unfortunately the beta-delayed left region contains emission data from excited state of nuclei
with $A=20,~21$ and the corresponding symbols lie practically on a vertical line in Fig. \ref{fig8} (a).
Therefore the predicting power of the $A^{1/3}$ dependence in this region cannot be checked and we hope
that future experimental data will confirm the slope rule (\ref{estim}).

On the other hand, it is known that the reduced decay width for spontaneous proton emission processes
is related to the fragmentation potential $V_{frag}=V_B-Q$. In order to check the validity of this law
we plotted in Fig. \ref{fig9} the logarithm of the monopole Coulomb-reduced width versus the fragmentation potential.
It turns out that the experimental data follow an approximate linear correlation 
\bea
&&\log_{10}\gamma_0^2=aV_{frag}+b~,
\nn
&&a=-0.015,~b= -1.246,~\sigma=0.509~,
\eea
with a common negative slope, as predicted by Refs. \cite{Del09,Del21}. 
Let us mention that the slope can be estimated by approximating the internal nuclear barrier with 
an inverted parabola with frequency $\hbar\omega_N$ \cite{Del21}
\bea
a=-\frac{\pi\log_{10}e}{2\hbar\omega_N}~.
\eea

\begin{figure}[ht] 
\begin{center} 
\includegraphics[width=9cm]{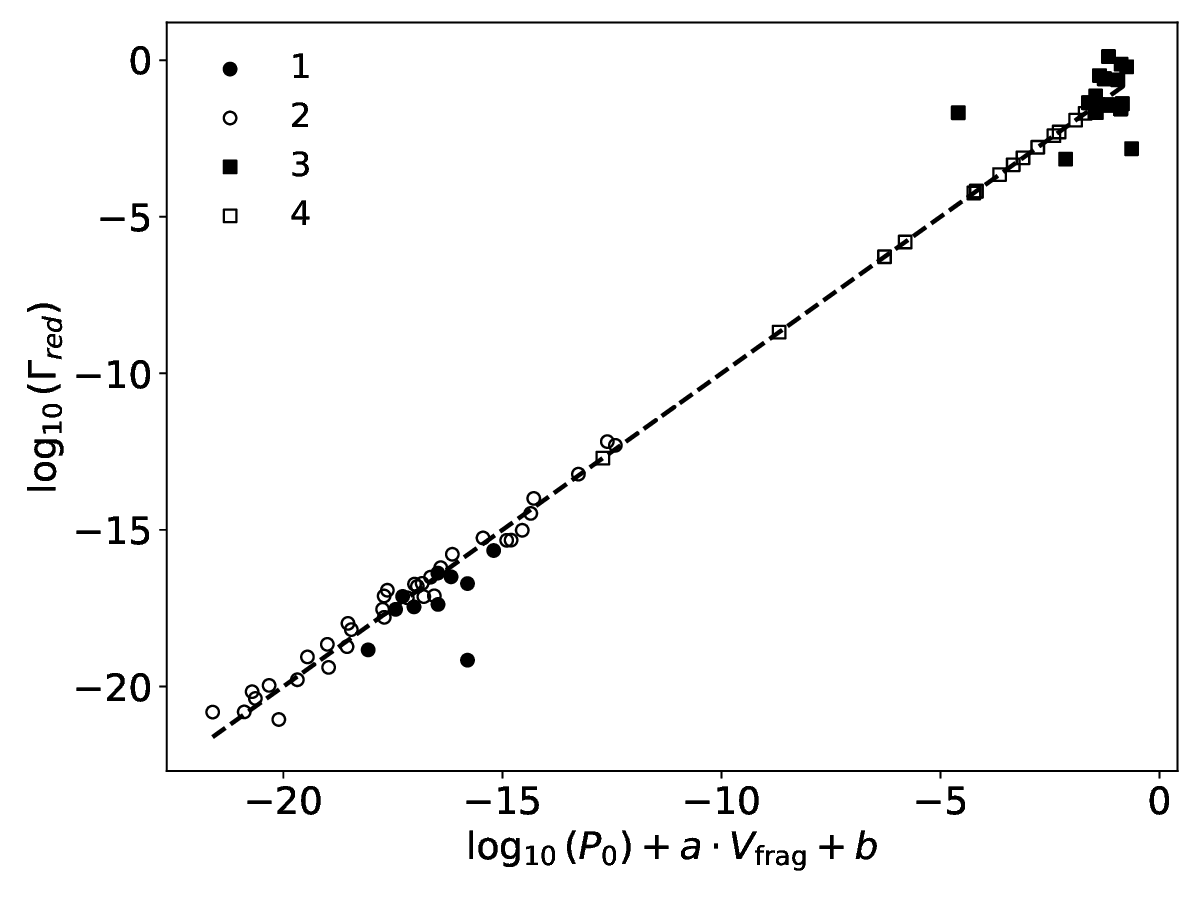}
\vspace{-5mm}
\caption{
Logarithm of the monopole Coulomb-reduced decay width (\ref{red}) versus $\log_{10} P_0+aV_{frag}+b$.
Different symbols denote the four zones in Table I.
}
\label{fig10}
\end{center} 
\end{figure}

Based on this conclusion we are able to use the following general dependence
\bea
\log_{10}\Gamma_{red}=\log_{10}P_0(\chi,\rho)+aV_{frag}+b~,
\eea
where $\rho=\kappa R_0$, $R_0=1.2A^{1/3}$,
providing a satisfactory representation for the monopole Coulomb-reduced decay width
of the available proton emission data, as can be seen in Fig. \ref{fig10}. The mentioned logarithmic 
rms error corresponds to a factor of three in the uncertainty predicting the absolute values
of the decay width. The symbols are given in Table I. 
The "fine details" induced by the angular momentum dependence of the internal proton wave function are not seen
in this 20 orders of magnitude scale. 
We also represented here together with measured decay widths the predictions for the emitters 
in the 4-th zone by open squares. 
Let us mention that decay widths with different angular momenta of Coulomb waves can be estimated according
to Eq. (\ref{red}).

\section{Conclusions}

\label{sec:concl} 
\setcounter{equation}{0} 
\renewcommand{\theequation}{4.\arabic{equation}} 

We analyzed the proton emission along proton drip line, by comparing spontaneous emission
occurring at nuclei with $A>100$ to beta-delayed proton emission from excited states in nuclei with $A<100$.
We described these processes in terms of the outgoing Coulomb-Hankel wave function, 
depending upon angular momentum $l$, Coulomb parameter $\chi$ and reduced radius $\rho$.
We analyzed the proton emission chart in ($l$, $\chi$, $\rho$) variables. It turns out that 
the decay width $\Gamma$ of a resonant state in a proton mean field given by 
the use of the continuity equation for Gamow states and phase shift analysis of real scattering states
are numerically very close, while the direct integration of the Schr\"odinger equation in the complex plane
gives slightly different results for $\log_{10}\Gamma>-7$.
We investigated the role of the centrifugal barrier induced by Coulomb waves and the dependence
of the monopole Coulomb-reduced decay width upon mass number and angular momentum,
predicted by single particle orbitals. The angular momentum splitting induced by nuclear orbital
is confirmed by spontaneous proton emission data and is predicted for beta-delayed proton emitters. 
We were able to obtain a generalized Geiger-Nuttal law linearly connecting the logarithm of the 
experimental monopole Coulomb-reduced decay width to the logarithm of the monopole Coulomb penetrability and 
fragmentation potential in terms of only two parameters for all proton emission processes.
The error in describing experimental decay widths by this dependence corresponds to a factor of three
for absolute values.
This law provides an useful tool to estimate decay widths of proton emitters along the proton drip line.

\begin{acknowledgments}
This work was supported by a grant of the Ministry of Research, Innovation and
Digitization, CNCS - UEFISCDI, project number PN-IV-P1-PCE-2023-0273, within PNCDI IV.
\end{acknowledgments}


\end{document}